\documentstyle[aps,twocolumn]{revtex}

\def\s#1{\vert#1\rangle}
\def\cs#1{\langle#1\vert}
\def\tr{{\rm tr}}

\begin{document}

\title{Quantum authentication and key distribution using catalysis}

\author{Jens G.\ Jensen and R\"udiger Schack}

\address{Department of Mathematics, Royal Holloway, University of London,
Egham, Surrey TW20 0EX, UK}

\date{13 June 2000}
\maketitle

\begin{abstract}
  Starting from Barnum's recent proposal to use entanglement and catalysis for
  quantum secure identification [{\tt quant-ph/9910072}], we describe a
  protocol for quantum authentication and authenticated quantum key
  distribution.  We argue that our scheme is secure even in the presence of an
  eavesdropper who has complete control over both classical and quantum
  channels. 
\end{abstract}


Since the publication of the BB84 protocol \cite{Bennett1984}, quantum
key distribution has developed into a well-understood application of
quantum mechanics to cryptography.  Typically, quantum key distribution
schemes depend either on an unjammable classical communication channel
or on authentication of the classical communication by classical
methods.  Comparatively little work has been done on the problem of
quantum authentication and authenticated quantum key distribution.

Some existing quantum authentication proposals are variations of the
BB84 protocol \cite{Crepeau1995,Huttner1996,Dusek1999a}.  These proposals
either require an unjammable classical channel, or authentication of the
classical communication using classical cryptographic methods
\cite{Dusek1999a}.  An early proposal \cite{Crepeau1995} uses quantum
oblivious transfer, which has since been shown to be insecure
\cite{Lo1997a}. Some recent proposals \cite{Barnum-9910,Zeng2000a,Zeng-0001c}
are based on entanglement.  A very interesting protocol
of this type is due to Howard Barnum \cite{Barnum-9910}.  In his
protocol, the parties use a shared entangled pair of particles as a
catalyst \cite{Jonathan1999a} to perform a quantum operation which would
be impossible without the catalyst.  In its original form, however,
Barnum's protocol has been shown to be insecure \cite{BK}.

In this paper, we describe a protocol derived from Barnum's protocol which
appears to be secure against a wide range of eavesdropping attacks.  In a
simplified version of our protocol, the two parties, Alice and Bob, initially
share $K$ particle pairs in an entangled state $\s{c}$ (the \textit{key} or
\textit{catalyst}).  Assume Alice wants to identify herself to Bob.  Bob then
prepares $K$ pairs of particles in an entangled state $\s{b}$ and sends one
particle from each pair to Alice (the {\it challenge}). 
It is possible \cite{P} to choose the states $\s{c}$ and
$\s{b}$ such that by using only local operations and classical
communication (LQCC), Alice and Bob can convert the four-particle state
$\s{b}\s{c}$ into the four-particle state $\s{c}\s{c}$, but by using only
LQCC, the two-particle state $\s{b}$ cannot be converted into the two-particle
state $\s{c}$ deterministically.  The state $\s{c}$ thus acts as a catalyst
\cite{Jonathan1999a} for the conversion of $\s{b}$ into $\s{c}$.

Using a different catalyst for each pair of challenge particles, Alice
and Bob perform LQCC to convert all $K$ challenge pairs to the state
$\s{c}$.  Bob now selects a number $K'$ of his challenge particles and
asks Alice to send back her corresponding challenge particles (her
\textit{response}).  For each of the $K'$ challenge pairs now in his
possession, Bob makes a projective measurement onto the state $\s{c}$.
An eavesdropper, Eve, pretending to be Alice, would not have had access
to the catalyst $\s{c}$, so Eve and Bob would not have been able to
convert all their challenge particles to the state $\s{c}$, and
therefore some of Bob's test measurements would fail.  Below we will
derive an upper bound $p_0$ for the probability $p$ that an eavesdropper
remains undetected in a single such measurement.  The overall
probability of not detecting an eavesdropper is bounded above by
$p_0^{K'}$ and can be made arbitrarily small by choosing $K'$ large
enough.

After a successful authentication, Alice and Bob share $2(K-K')$ catalyst
pairs, since the protocol requires that they destroy the catalyst pairs
used in the conversion of the $K'$ tested challenge pairs.  If $K>2K'$, they
now share more key particles than before.  Our authentication protocol thus
also provides authenticated quantum key distribution.

The simplified version of our protocol just given is not secure.  Below,
we first describe a full version of the protocol, and then we discuss a
number of eavesdropping attacks against it which we believe are the most
powerful such attacks.  We will argue that our protocol is secure even
in the presence of an eavesdropper with full control over both classical
and quantum communication channels; we do not, however, give a full security 
proof.  In our analysis, we assume that all
quantum operations are error-free and that the quantum channel is
noiseless.

\smallskip\noindent{\bf Choice of states.}  Consider bipartite states
$\s{b}=\sum_{k=1}^n\sqrt{b_k}\s{k}\s{k}$ and
$\s{c}=\sum_{k=1}^n\sqrt{c_k}\s{k}\s{k}$, where the states $\s{k}$ are
orthonormal basis states for one particle.  If $b_1\ge\cdots\ge b_n$ and
$c_1\ge\cdots\ge c_n$, then $b_k$ and $c_k$ are  called the ordered Schmidt
coefficients of the states $\s{b}$ and $\s{c}$.  The state $\s{b}$ can be
converted deterministically into $\s{c}$ using only LQCC iff the ordered
Schmidt coefficients of the target state $\s{c}$ {\it majorize\/}
those of the initial state $\s{b}$ \cite{Nielsen1999}, i.e., iff
$\forall{}k:\sum_{i=1}^k c_i\geq\sum_{i=1}^k b_i$ with equality for
$k=n$.  Otherwise, only a probabilistic conversion is
possible \cite{Jonathan1999b,Vidal1999b}.

States with the properties required for our protocol exist for $n=5$ \cite{P}.
For $n\le4$, the protocol needs to be modified to use probabilistic
entanglement-assisted conversion \cite{Jonathan1999b}.  Our choice of Schmidt
coefficients for $\s{b}$ and $\s{c}$ is $b_1=b_2=0.31$, $b_3=0.30$,
$b_4=b_5=0.04$, $c_1=0.48$, $c_2=0.24$, $c_3=c_4=0.14$, $c_5=0$.  With
this choice, the conversion of $\s{b}$ into $\s{c}$ can be done only with
probability $P(b\to c)\simeq0.572$, but the ordered Schmidt coefficients of
the tensor-product state $\s{c}\s{c}$ majorize those of the state
$\s{b}\s{c}$, so the latter can be converted into the former
deterministically.

Even though the exact conversion $\s{b}\to\s{c}$ can only be done with
probability 0.572, it is possible to convert $\s{b}$ to pure or mixed states
$\rho$ close to $\s{c}$ with much higher probability (we say that $\rho$ is
close to $\s{c}$ if the fidelity $F=\langle c|\rho|c\rangle$  is close to
1).  By applying a theorem given in Ref.~\cite{Vidal-9910}, it can be seen that the
average fidelity for the conversion $\s{b}\to\s{c}$ is bounded above
as
\begin{equation}
\langle c|\rho|c\rangle \le p_0\simeq0.9907 \;,
\label{eqfid}
\end{equation}
where $\rho$ is now the average state resulting from the conversion. 
The theorem also shows that the maximum average
fidelity $\bar F=p_0$ is achieved by a pure state $\s{\xi_c}$ to which $\s{b}$
can be converted deterministically.

\smallskip

%% These used to be labels; they refer to the steps in the protocol below
\def\enkrprep{1}
\def\enkrconv{2}
\def\enkrauth{3}
\def\enkrfail{4}
\def\enkrlabel{5}

\noindent\textbf{Overview of the full protocol.}
The main difference between the simplified version of the authentication
protocol given above and the full version is that the latter is
symmetric.  In an {\it authentication round}, Alice and Bob each
establish the identity of the other part.

One round of the protocol consists of Alice and Bob each preparing $K$
particle pairs in state $\s{b}$.  Bob sends one particle of each of his pairs
to Alice; for these pairs, Alice is called the {\it prover\/} and Bob the {\it
  verifier}.  Likewise, Alice sends one particle of each of her pairs to Bob;
for these pairs, she is the prover and he the verifier.  Using a different
catalyst for each pair, Alice and Bob now convert each of the $\s{b}$ states
to a $\s{c}$ state.  Each of the two asks the other to send back $K'$
($K'<K/2$) of the
new particles for testing; they abort the protocol if they detect any particle
pair not in the $\s{c}$ state.

Eve, who does not initially share any entanglement with Alice and Bob,
cannot impersonate one of them to the other.  For a successful attack,
Eve must therefore first obtain shared entanglement with Alice and Bob.
Below, after describing the protocol in detail, we discuss its security
against a number of attacks, where Eve has full control over both
the quantum and classical communication channels (such attacks are
called ``man-in-the-middle'' attacks).

\smallskip
\noindent\textbf{The key.}
Before the first authentication round, we shall assume that Alice and Bob
share $2K$ particle pairs prepared in the state $\s{c}$: these are the
catalysts, and together they form the key.  With each successful
authentication round, the number of key pairs increases.  In each round, the
key particles used are labeled $\gamma_A^i$ and $\gamma_B^i$, respectively,
where $i=1,\dots,2K$, and the state of each pair $\gamma_A^i\gamma_B^i$ is
$\s{c}$.

\smallskip
\noindent\textbf{Detailed description.} An authentication round consists of
the following steps.

\noindent\enkrprep.   Bob prepares $K$ particle pairs $\beta_A^i\beta_B^i$
  in state $\s{b}$, where $i$ is \textit{odd}, and sends $\beta_A^i$ to Alice.
  These are Bob's \textit{challenges}.  Likewise, Alice prepares $K$ pairs
  $\beta_A^i\beta_B^i$ in state $\s{b}$ for $i$ even, and sends $\beta_B^i$
  to Bob.  Thus, for odd indices, Bob will be the verifier; Alice will be the
  verifier for even indices.

\noindent\enkrconv.   For each $i$, Alice and Bob perform the deterministic
catalysis conversion $\s{b}\s{c}\longrightarrow\s{c}\s{c}$, where Alice
performs local operations on her particles $\gamma_A^i$ and $\beta_A^i$ and
Bob performs local operations on his particles $\gamma_B^i$ and $\beta_B^i$
\cite{Nielsen1999}. We can \cite{Lo-9707} and do require that only the
verifier performs both unitary transformations and generalized measurements;
the prover performs only unitary transformations depending on the result of
the verifier's measurements, which are communicated classically.

\noindent\enkrauth.  Alice picks randomly a subset
$Q_A\subseteq\{2,4,\dots,2K\}$ of size $K'$ of particles for which she
is the verifier, and Bob does likewise for a subset
$Q_B\subseteq\{1,3,\dots,2K-1\}$ of size $K'$ for which he is the
verifier.  Bob as verifier now asks Alice to send back her
\textit{response} $\beta_A^i$ for some $i\in{}Q_B$.  Bob measures the
projector $|c\rangle\langle c|$ on the particle pair
$\beta_A^i\beta_B^i$.  If the measurement fails, he aborts the protocol.
Then Alice becomes the verifier, asks Bob to send $\beta_B^i$ for some
$i\in{}Q_A$ and tests it likewise.  They continue taking turns as prover
and verifier until they have exhausted the sets $Q_A$ and $Q_B$.  At the
end of this step, they discard the catalysts $\gamma_A^i\gamma_B^i$ for
$i\in Q_A\cup Q_B$.

\noindent\enkrfail.  The authentication fails if
any of the projective measurements in the previous step fails, 
or if Alice or Bob
receive more than $K'$ requests to send back challenge particles.

\noindent\enkrlabel.  If the authentication round succeeds, Alice and Bob 
are left with $2(K-K')$ pairs $\gamma^i_A\gamma^i_B$ and $2(K-K')$ pairs
$\beta^i_A\beta^i_B$, i.e., they now have $2K-4K'$ additional pairs in the
catalyst state $\s{c}$. The $2K+n(2K-4K')$ they share after the $n$th
successful round are now renamed $\gamma^j_A\gamma^j_B$ in random order, i.e.,
with the indices $j$ permuted using a pseudo-random number generator.

\smallskip
\noindent\textbf{Remark.} If the authentication fails, the parties discard
all particles used till that point, including both the original key and
all new key pairs generated.  In this case, Alice and Bob have to start
again with a new key.  Therefore, in practice they should initially
share several sets of $2K$ key pairs.

\smallskip
\noindent\textbf{Security and attacks.}
We now dicuss the security of our protocol against a number of attacks.  
We start
with two simple attacks, impersonation and denial of service, and then move on
to more powerful ``man-in-the-middle'' attacks.

\smallskip \textit{Impersonation}.  Suppose that Alice is not present and Eve
tries to persuade Bob that she is Alice.  When Bob sends out a challenge
particle, Eve intercepts it.  We therefore label it $\beta_E$ rather than
$\beta_A$, omitting the index $i$ for clarity.  Eve must now perform local
operations on $\beta_E$ such that a later measurement by Bob on the pair
$\beta_E\beta_B$ will fail with the smallest possible probability.  If $\rho$
is the average state of the pair $\beta_E\beta_B$ resulting from Eve's and
Bob's operations, then the probability that Bob's measurement succeeds is
given by the fidelity $\langle c|\rho|c\rangle$.  Since Eve does not
have the catalyst particle $\gamma_A$ paired with the particle $\gamma_B$ that
Bob will use, the conversion is not assisted by any entanglement.  The
fidelity $\langle c|\rho|c\rangle$ is therefore bounded above by $p_0<1$ [see
Eq.~(\ref{eqfid})].  Since in one authentication round, Bob makes $K'$ such
measurements, the probability of not detecting Eve is bounded above by
$p_0^{K'}$, which can be made arbitrarily small by choosing $K'$ large enough.

\smallskip \textit{Denial of service}.  In this type of attack, Eve
deliberately causes the authentication round to fail, and hence causes one
party to discard all key particles.  Although our protocol in its present form
is particularly vulnerable to this kind of attack, this is not an essential
weakness since an attacker who controls both quantum and classical
communication can always prevent successful authentications between the
legitimate parties.

\smallskip \textit{Man in the middle}.  We now look at stronger attacks in
which Eve tries to obtain key material which she could then use, e.g., in a
later impersonation attack.  Eve's goal is to share pairs of
particles in the catalyst state $\s{c}$ with Alice and/or Bob.  For instance,
if she succeeds in obtaining a large amount of key material shared with Bob,
she will be able to authenticate herself to Bob without Alice being present.
Eve's ability to obtain key material is limited by the fact that if her
presence is detected in a single measurement, all the previously obtained
key material she shares with the verifier who performed that measurement 
will become worthless. 

We will distinguish between two kinds of attacks.  In a {\it type I attack},
Eve does not intercept the challenge particle when it is sent from the
verifier to the prover.  In a {\it type II attack}, she intercepts the
challenge particle and sends another particle on to the prover.  Since the
protocol is symmetric, we will  assume in the following that Alice is
the prover and Bob the verifier.

\smallskip \textit{Type I attack}.  By definition, in a type
I attack, Bob sends the challenge particle $\beta^i_A$ to Alice without Eve
interfering.  Assume now that Bob sends out a request for a response particle.
Eve has three options.  In option 1, she passes the request on to Alice, then
she passes Alice's response particle $\beta^i_A$ on to Bob.  Eve's presence will
not be detected, but she does not obtain any key material either.  In option 2,
Eve passes the request on to Alice, then intercepts Alice's particle and sends
another particle on to Bob.  Eve does not gain anything, because both Alice and
Bob are going to discard their respective particles.  In addition, Eve risks
detection with nonzero probability.

Option 3 is the interesting one.  Here, Eve does not pass Bob's request on to
Alice.  Instead she prepares a pair of particles $\alpha_E$ and $\alpha_B$ in
a state of her choice and sends $\alpha_B$ to Bob.  Then she asks Alice to
send back the particle $\beta^{i+2}_A$, which is the next one for which Bob is
the verifier.  Since the pair $\beta^{i+2}_A\beta^{i+2}_B$ is in the state
$\s{c}$, Eve now shares a perfect catalyst pair with Bob (assuming that
$i+2\not\in Q_B$).  Bob's measurement on the pair $\alpha_B\beta^i_B$,
however, is going to detect her with a probability not less than $1-p_0$.  In
the case that Bob's measurement does not detect her, we assume for our
security analysis that, after the measurement, the pair $\alpha_E\beta^i_A$
shared between Alice and Eve is in state $\s{c}$, which is probably too strong
an assumption. There is an additional risk of detection for Eve in the next
authentication round since, when Alice and Bob relabel their particles in step
\enkrlabel{} of the protocol, there will be a $j$ such that $\gamma^j_A$
is not entangled with $\gamma^j_B$.

Even if Bob does not ask for a response particle, Eve may still send a request
to Alice, so that again she obtains a perfect catalyst pair with Bob. However,
since Alice will abort the protocol if she receives more than $K'$ requests to
send back a response particle, Eve cannot request a particle from her without
also at some time during the round sending a corresponding response particle
to Bob.  Therefore, Eve cannot avoid being detected with a probability of at
least $1-p_0$ for each key particle she obtains in this way.

\smallskip \textit{Type II attack}.  
We now assume that Eve intercepts the challenge particle $\beta_A$ sent out by
Bob.  As before, because Eve now owns that particle, we will label it
$\beta_E$.  The pair $\beta_E\beta_B$ is in state $\s{b}$.  Eve then
prepares two particles $\alpha_A\alpha_E$ in a state $\s{a}$ of her
choice, keeps $\alpha_E$ and sends $\alpha_A$ to Alice. 

Unaware of Eve's presence, Bob now goes through the catalysis protocol with
his particles $\beta_B$ and $\gamma_B$, where $\gamma_B$ is entangled with
Alice's particle $\gamma_A$.
Bob sends out the results of
his generalized measurements, which Eve intercepts.
Bob's two particles $\gamma_B$ and $\beta_B$ are now in the state
\begin{equation}
\rho_{\gamma_B\beta_B} =
\tr_{\gamma_A\beta_A}(\rho_{\gamma_A\gamma_B}\otimes\rho_{\beta_A\beta_B})
=\tr_{\gamma_A\beta_A}(\s{cc}\langle cc|) \;.
\label{eqtraces}
\end{equation}
This state is independent of Alice's and Eve's actions and has no entanglement
between the two particles.

At this point, there are three different cases.  In the first case, Bob does
not request a response particle; Eve thus does not risk being detected.
She now shares entangled states with both Alice and Bob.  She can perform
arbitrary unitary or nonunitary local operations on her particles $\alpha_E$
and $\beta_E$, and she can send fake measurement information to Alice in order
to influence Alice's unitary operations.  For our security analysis, we assume
that this enables her to bring both pairs $\alpha_A\alpha_E$ and
$\beta_E\beta_B$ into the catalyst state $\s{c}$, although it follows from the
analysis of case 2 below that she cannot reach this goal completely.  Eve may
also ask Alice to send particle $\alpha_A$ back to her, but generally, Eve
will not gain anything from this.

In the second case, Bob requests a response particle, and Eve sends him her
particle $\beta_E$.  We will now show that the fidelity between the target
state $\s{c}$ and the state $\rho_{\beta_E\beta_B}$ on which Bob performs his
measurement is bounded above by $\cs{c}\rho_{\beta_E\beta_B}\s{c}\le p_0$,
which implies that Bob's measurement fails with probability $\ge1-p_0$.

The reason is that even if Bob collaborated with Eve on maximizing the
fidelity, they could only use LQCC in the conversion; it would not be assisted
by any entanglement.  Since Alice performs only unitary transformations, but no
measurements, on her particles $\gamma_A$ and $\alpha_A$, no
entanglement is created between $\alpha_E$ and $\gamma_B$, which could assist
Eve and Bob in their task.

As in the first case, for our security analysis we will assume that if Eve
remains undetected, she shares pairs in the catalyst state with both Alice and
Bob.  Eve can get close to this goal by performing a type I attack against
Alice leading to a perfect catalyst pair shared with Bob.  Eve can do this
because she has not passed Bob's earlier request on to Alice.

In the third case, Bob also requests a response particle, but this time Eve
passes his request on to Alice and intercepts Alice's response $\alpha_A$.  Eve
then performs arbitrary operations on the three particles now in her
possession, $\alpha_A$, $\alpha_E$ and $\beta_E$.  Then she sends one particle
on to Bob.  We label this particle $\tilde\beta_E$. 

We now assume that Eve does not use any entanglement to assist her in the
conversion of the $\beta$ particles, which means that the fidelity between the
target state $\s{c}$ and the state $\rho_{\tilde\beta_E\beta_B}$ on which Bob
performs his measurement is bounded above by
$\cs{c}\rho_{\tilde\beta_E\beta_B}\s{c}\le p_0$. This implies again that Bob's
measurement fails with probability $\ge1-p_0$.

The above assumption is rather strong, but partially justified by the
fact that there is a conflict of interest for Eve: if Bob does not request a
response particle, Eve wants the $\alpha$ particles to be in the pure $\s{c}$
state, in which case they are not entangled with any other particle. For a
full analysis of this conflict of interest, one needs to analyse the set of
unitary transformations Alice is allowed to perform under the protocol.

Unlike the first and second cases, if Bob's measurement does not fail, Eve
will not share entanglement with either Alice or Bob, since they discard their
respective particles.

To evaluate the overall security of the protocol against a type II attack, we
now assume that Eve attacks $L$ particle pairs.  Since Alice and Bob check a
random fraction $K'/K$ of these pairs, the probability that Eve remains 
undetected is
approximately bounded above by $p_0^{LK'/K}$---the bound becomes exact in the
limit of large $K$ and $K'$.  If Eve is not detected, the fraction $e$ of key
pairs she shares with Alice and Bob is not greater than $L/K$.  The probability
$p(e)$ that Eve obtains a fraction $e$ undetected is therefore bounded above
by $p_0^{eK'}$.  The security of the protocol against a type II attack then
follows from the fact that, for any $e>0$, Alice and Bob can make $p(e)$
arbitrarily small by choosing $K$ and $K'$ sufficiently large.

Similarly, the protocol is secure against a type I attack because the
probability that Eve remains undetected in a type I attack against $L$ particle
pairs is bounded above by $p_0^L$.

\smallskip
\noindent\textbf{Conclusions and outlook.} 
The quantum authentication protocol described above appears to be secure even
in the presence of an eavesdropper who has complete control over both
classical and quantum communication channels at all times.  Our protocol does
not rely on classical cryptography.  Furthermore, the security of the protocol
does not depend on keeping classical information secret, including information
about quantum states: all parties, including the eavesdropper, have full
information about all aspects of the protocol.  In each authentication round,
additional quantum key particles are distributed securely.  Combined with
entanglement purification and privacy amplification techniques
\cite{Deutsch1996}, our protocol therefore also provides authenticated quantum
key distribution.

There is a number of important open questions which we plan to address in the
future.  Most importantly, we need to analyse the protocol in the presence of
noise and for more subtle eavesdropping attempts such as coherent attacks, or
an attack in which
Eve partially entangles the challenge with an ancillary particle. 
Furthermore, there is scope for
improving the protocol in several respects.  For instance, the parties should
not have to discard all key pairs if a single measurement fails.  It should 
also
be possible to find states with a lower fidelity bound $p_0$, e.g., by going
to a higher-dimensional Hilbert space.

\smallskip
\noindent\textbf{Acknowledgments.}  
The authors would like to thank Howard Barnum, Todd Brun and Martin
Plenio for very helpful discussions. Thanks also to Daniel Gottesman and
Norbert L\"utkenhaus for pointing out problems with previous versions 
of the protocol.  This work
was supported by the UK Engineering and Physical Sciences Research
Council (EPSRC).

\end{document}